\documentclass[a4paper,11pt]{article}
\usepackage{jinstpub} 
\usepackage{lineno}
\usepackage{placeins}
\usepackage{subcaption}

\title{\boldmath Wavelength-tunable femtosecond pulsed laser for the characterisation of solid-state sensors}

\author[a,e]{E. Ejopu \footnote{Equal Contribution: Enoch and Nawal contributed equally to this work.}}
\author[a]{, O.A. de A. Francisco}
\author[b,d]{, N. Al Amairi}
\author[a]{, H. Li}
\author[a,c]{, M. Gersabeck}
\author[a]{, A. Oh}
\author[a]{, C. Parkes}
\author[b]{, P. Parkinson}
\author[a]{, F.M. Sanchez}
\author[b]{and C. Smith}
\affiliation[a]{Department of Physics and Astronomy, The University of Manchester,\\ 
Schuster Building, Manchester M13 9PL, United Kingdom}
\affiliation[b]{Photon Science Institute, The University of Manchester,\\ 
Alan Turing Building, Oxford Rd, Manchester M13 9PY, United Kingdom}
\affiliation[c]{Physikalisches Institut, Albert-Ludwigs-Universit\"at Freiburg,\\
Hermann-Herder-Stra\ss{}e 3b, 79085 Freiburg im Breisgau, Germany}
\affiliation[d]{Department of Applied Sciences, Physics Section, University of Technology and Applied Sciences,\\
P.O Box 74, Al-Khuwair, PC 133, Muscat, Oman}
\affiliation[e]{Department of Physics, University of Liverpool,\\
The Oliver Lodge, University of Liverpool, Oxford St, Liverpool L69 7ZE, United Kingdom}
\emailAdd{oscar.augusto@manchester.ac.uk, eejopu@gmail.com}

\abstract{The Two-Photon Absorption (TPA) process provides an excellent method for the characterisation of solid-state sensors due to its intrinsic 3D resolution for the excitation of electron-hole pairs. This paper presents the commissioning of a system composed of a Light Conversion PHAROS laser read out with an ORPHEUS optical parametric amplifier with tuneable output wavelength across $310-16000$~nm. This is one of the few facilities suitable for particle physics detector applications with a very wide wavelength range and ultra short pulses. The voxel characterisation shows a Rayleigh length of $12.20 \pm 0.47$ $~\upmu$m and the beam radius at the waist of $1.53 \pm 0.04$~$\upmu$m. Initial measurements with silicon and diamond samples are described. The apparatus is characterised with both devices, demonstrating the expected quadratic dependency with the energy per pulse, and different models are presented to explain the signal generated due to reflection. The signal generated in the silicon diode is in good agreement with one dimension voxel model predictions.}


\arxivnumber{2509.01546} 

\begin{document}
\maketitle
\flushbottom

\section{Introduction}
\begin{figure}[htbp]
    \centering    
    \includegraphics[width=0.99\linewidth]{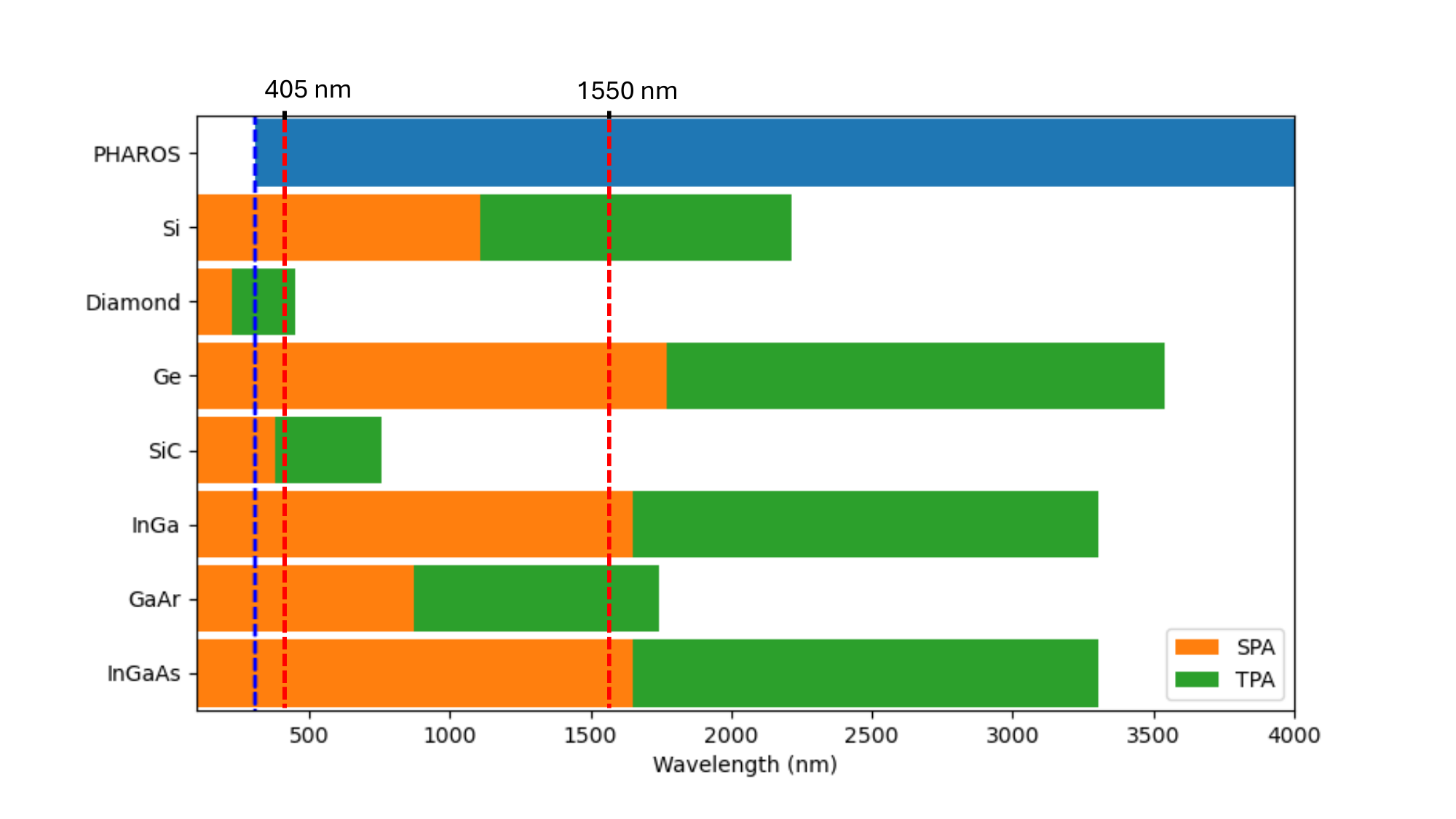}
    \caption{Comparison of the PHAROS + ORPHEUS wavelength range for single-photon absorption (SPA) and two-photon absorption processes for several solid-state sensors (non-irradiated). The dashed red lines show the laser wavelengths used in this study. The dashed blue line shows the minimum limit of the laser, with the maximum being at 16000~nm.}
    \label{fig:pharosrangeandsensors}
\end{figure}

\FloatBarrier
Solid-state detectors are used in a wide range of fields, these include applications of particular interest to the authors in fundamental particle physics research at the vertex detectors and tracking systems at the LHC~\cite{VELOU1TDR} and in research on medical imaging technologies such as positron emission tomography (PET)~\cite{PET}. The characterisation of new designs and substrates help to guide the design choices made during the research and development phase.
The transient current technique with two photon absorption (TPA) has been used for the characterization of different materials due to its intrinsic 3D resolution~\cite{pape2024characterisation, NIKHEFTPA, MoritzTheses, TPAGordana}. In TPA, a highly localised volume of charge near the focal point, referred to as a ``voxel'', is used in the characterisation of sensors. This work presents a set-up capable of testing a wide range of devices as indicated in Fig.~\ref{fig:pharosrangeandsensors} including silicon, diamond, SiC, InGa, Ge and more. At this stage, the initial results using a silicon diode and a planar diamond sample are described, thus demonstrating its potential to characterise different mediums. Furthermore, additional measurements with diamond and SiC samples have been performed, but they will be described in another article.

This paper demonstrates the potential of a single TPA-TCT setup to characterise several solid-state sensors\cite{Table_setup1, Table_setup2}. A tunable femtosecond source is used to cover a broad wavelength range (310 nm to 16,000 nm). 
The expected quadratic dependence on energy per pulse of the laser expected for the two photon absorption process is demonstrated with silicon and diamond samples. The   voxel characterisation using this system is then given with silicon samples, and it also includes the comparison with the modelling in the presence of reflections from a metallised back surface.

Fig.~\ref{fig:pharosrangeandsensors} illustrates the flexibility of the system covering the TPA wavelength ranges for several sensor substrates. With the wide wavelength range and the ability to perform an automated scan over several parameters simultaneously, including laser power, applied voltage, energy per pulse and the three spatial dimensions, this system provides excellent capabilities to characterise solid-state detectors.

\FloatBarrier

\section{Experimental setup}
\begin{figure}[h!]
    \centering
    \includegraphics[width=0.9\linewidth]{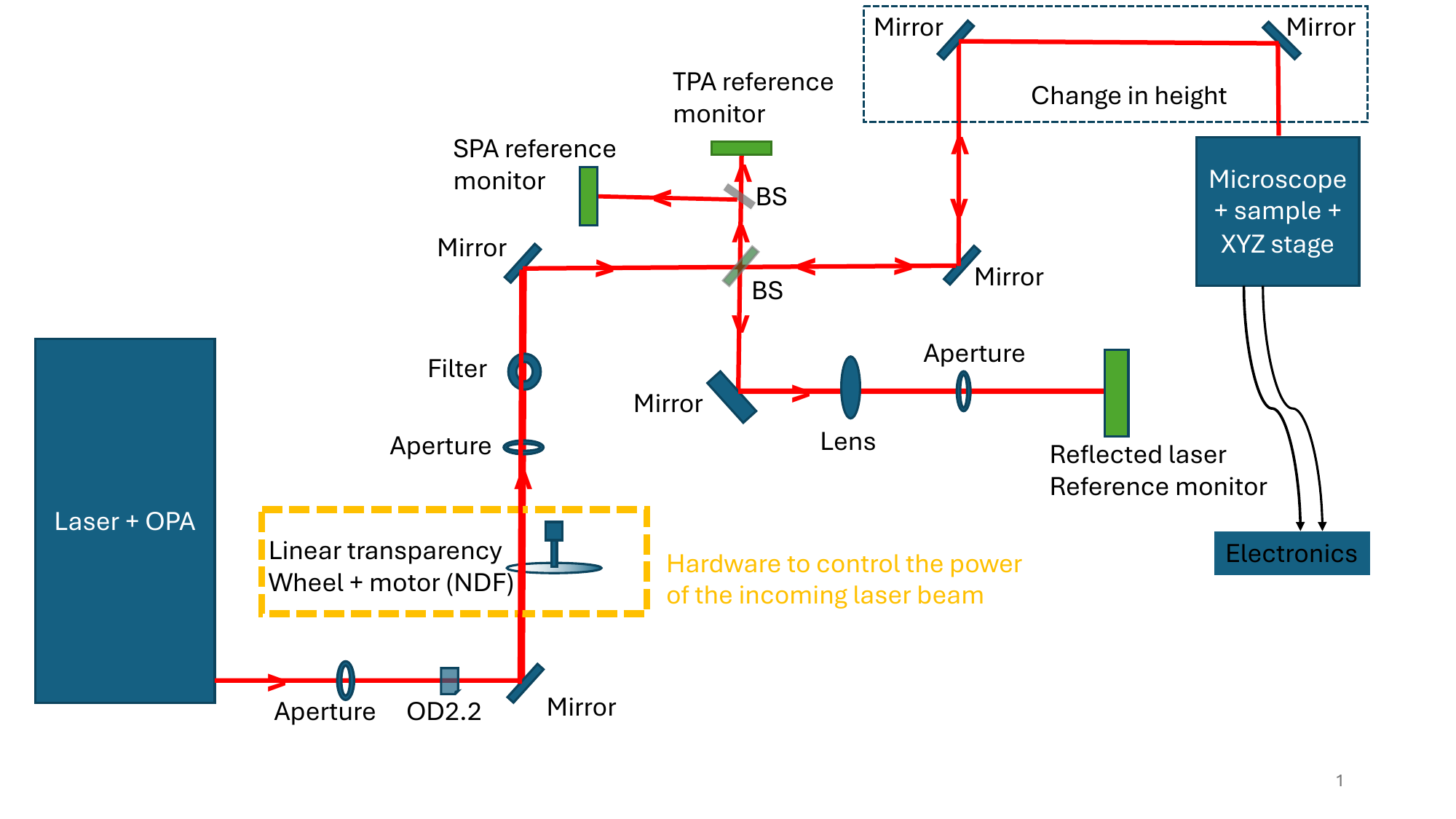}
    \caption{Schematic of the experimental set-up indicating optical components and the location of the monitors (SPA, TPA and reflected). BS refers to the Beam Splitter, while OD denotes the Optical Density filter.}
    \label{fig:Set_Up}
\end{figure}

A schematic of the TPA setup used is shown in Fig.~\ref{fig:Set_Up}. A Yb:YAG pump laser with 260~fs pulse duration and 200~$\upmu$J pulse energy\footnote{model PHAROS from UAB Light Conversion} is operated at up to 100~kHz repetition rate. Half of the output energy is used to pump an optical parametric amplifier (OPA), model ORPHEUS-HP from Light Conversion, which yields a tuneable output wavelength from 300~nm to 16000~nm, corresponding to a photon energy of 3.757~eV to 0.077~eV. Across this range the pulse duration varies from 170~fs to 200~fs, with typical pulse energies of several $\upmu$J. Silicon and diamond samples are characterised with different wavelengths, 1550~nm and 405~nm respectively. Since the fundamental wavelength of our laser is 1030 nm, these wavelengths (1550~nm and 405~nm) cannot be generated through simple harmonic generation. Therefore, an optical parametric amplifier is utilized.

The laser and OPA are situated on an optical table and the output beam is routed into an optical microscope.
The energy per pulse is controlled by a motorized variable reflective neutral density filter (NDF) wheel (continuous or discrete depending on the run) and additional fixed attenuation filters depending on the required laser intensity. 

A portion of the beam is separated and then split using two beam splitters, after which it is monitored using photodiodes. When the laser is operated at 1550~nm wavelength a germanium photodiode\footnote{Thorlabs model PDA30B2} tracks the beam intensity based on a single photon process, and a silicon photodiode\footnote{Thorlabs model PDA-100A-EC} placed at the focus of a 50~mm focal length lens tracks the variations based on the TPA process (same process as the device under test (DUT)). When operated at 405~nm, the same silicon photodiode is used without the focusing lens to track the single photon process. 

The laser is focused on the DUT with a microscope objective\footnote{Numerical Aperture, NA=0.5, 20x, Zeiss model 1156-521}. In optics, NA is a dimensionless parameter that provides a quantitative measure of the system's capacity to accept or emit light within a defined range of angles. The electrical signal induced in the DUT is amplified by a Cividec C2-TCT amplifier which includes a bias-tee for the supply of high voltage via a Keithley 2410 power supply. Reflected laser light from the DUT returns and is reflects from the beam splitter, from which it is focused onto a third photodiode\footnote{Thorlabs model PDA10CF-EC or model PDA36A-EC for operation at 1550~nm or 405~nm respectively}, allowing for a separate determination of the energy per pulse. The output signals from the photodiodes and the amplifier are then read by a Rohde \& Schwarz RTO1014 oscilloscope\footnote{1~GHz bandwidth, 10GS/s} with a sampling interval of 0.1~ns, triggered directly by the pump laser electronics. The control, monitoring and data acquisition were implemented using MATLAB\cite{MATLAB}. The offline analysis was performed using MATLAB or python, numpy, scipy and matplotlib libraries\cite{2020SciPy,pythonumpy,pythonmatplotlib}.

Two distinct silicon devices and one diamond device were investigated using the set-up. Silicon and diamond are used to demonstrate that the laser is wavelength tunable. The first silicon device is a silicon PIN diode (CIS diode, model CIS16-FZ-21) obtained from a FZ wafer~\cite{roeder2002rd50}, designed and fabricated at the Forschungsinstitut f\"{u}r Mikrosensorik (CiS) Research Institute for Micro Sensors GmbH. It has an active area of 5.0 $\times$ 5.0 mm$^{2}$ and a thickness of approximately 160~$\upmu$m in the active region. Openings in the back surface metallisation facilitate TCT and TPA measurements. This device was used for calibration and the measurement results compared with simulation. The second silicon device is a PIN diode fabricated at the Centro Nacional de Microelectr\'onica (CNM) in Barcelona. This has full back surface metallisation, thus allowing the test of the performance of  reflection modelling with the system. The CNM diode has an active area of 5.3 $\times$ 5.3~mm$^{2}$ and a thickness of 275~$\upmu$m~\cite{villegas2023gain}. The diamond device is a single-crystalline chemical vapour deposition (CVD) diamond sample from ElementSix~\cite{ElementSix2014}, measuring 4~mm $\times$ 4~mm  and a thickness of  500~$\upmu$m, with a partial Cr-Au surface metallisation on top and bottom sides.
\FloatBarrier

\section{Energy-Dependent Characterisation of Two-Photon Absorption}
\begin{figure}[h!]
    \centering
    \includegraphics[width=0.98\linewidth]{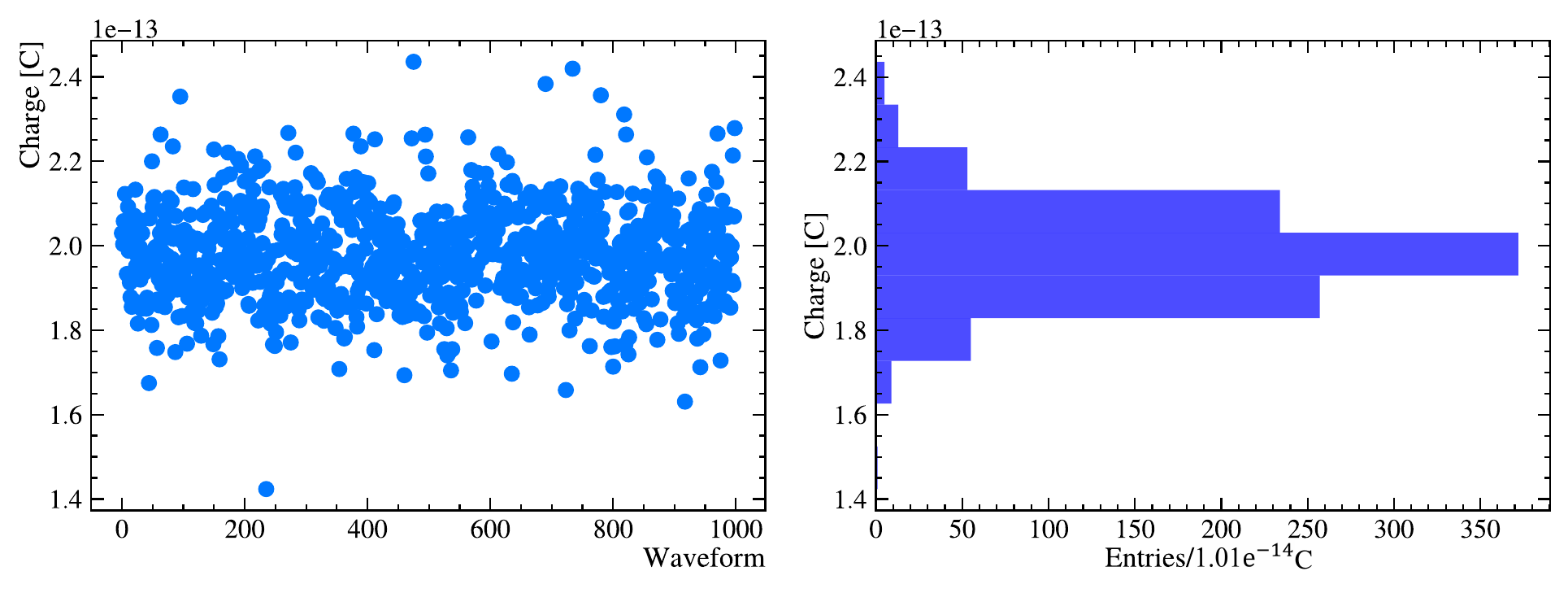}
    \caption{Charge measured in a CIS diode for 1000 pulses (\textit{left}) with a histogram showing the projection of the charge distribution on (\textit{right}). The variation is about 13\% of the mean charge.}
    \label{fig:charge_variation}
\end{figure}

\FloatBarrier
The variation of energy per pulse of the system as a function of time is illustrated in Fig.~\ref{fig:charge_variation}. The spread across 1000 single waveforms has a standard deviation of 13\% of the mean value. Hence, the standard deviation of the photon density is estimated to be around 6\% of the mean density (quadratic dependency). To minimize the impact of such variations, the signals are normalized by the maximum amplitude of the signal or the integrated charge observed in the SPA monitor. 

The energy-dependent characterisation involves two main steps, namely power calibration, and energy scan measurement. During the power calibration measurements, a power meter\footnote{Thorlabs PM100USB with S120C and S122C sensors} is placed in the DUT position, but in close proximity to the microscope objective (out-of-focus), the pulse repetition rate is set to 100 kHz and the motorized variable neutral density filter (NDF) is tuned to allow maximum laser power which is a position of maximum transparency. The NDF is then rotated from this position of higher power to a position of lower power while simultaneously recording the power meter and the maximum single-photon absorption (SPA) reference monitor readings. The energy per pulse calibration is obtained by a linear fit of the power meter reading against the maximum SPA amplitude. 

The effect of change in pulse frequency on the DUT waveforms is also verified. To do this, the laser frequency was varied from 200~Hz to 100~kHz while storing 100 waveforms. Taking into account the potential variations due to the pulse energy stability, no significant variation of the pulse shape with frequency was seen. 

\begin{figure}[htb]%
    \centering
    \begin{subfigure}{0.49\linewidth}
    \includegraphics[width=\linewidth]{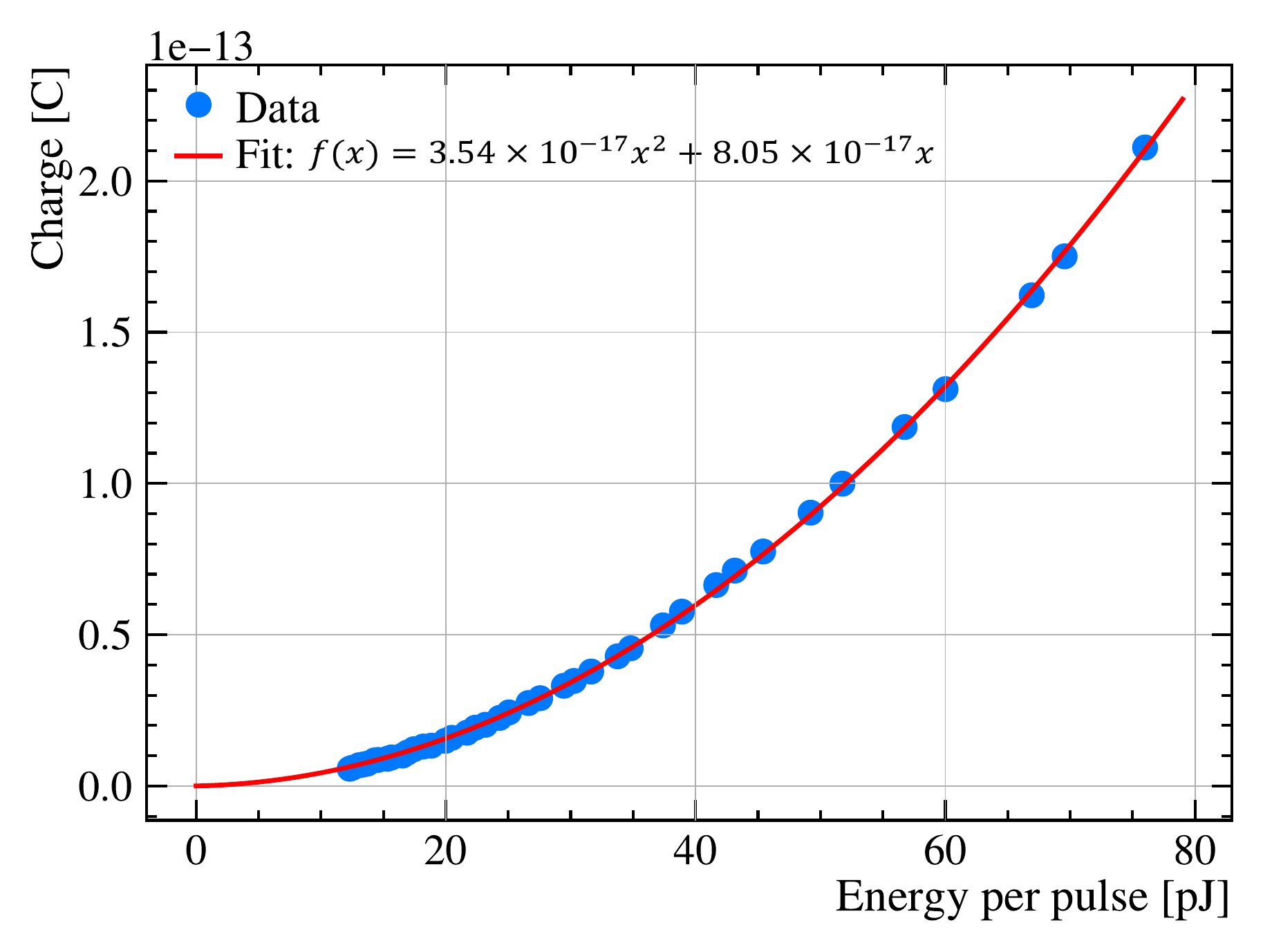}
 \caption{Charge vs energy per pulse for the silicon CIS diode sample at a laser wavelength of 1550~nm. A quadratic fit to the data points is shown.} \label{fig:DiodeChargeDepthScan}
  \end{subfigure}
  \hspace*{\fill}
    \begin{subfigure}{0.49\linewidth}
    \includegraphics[width=\linewidth]{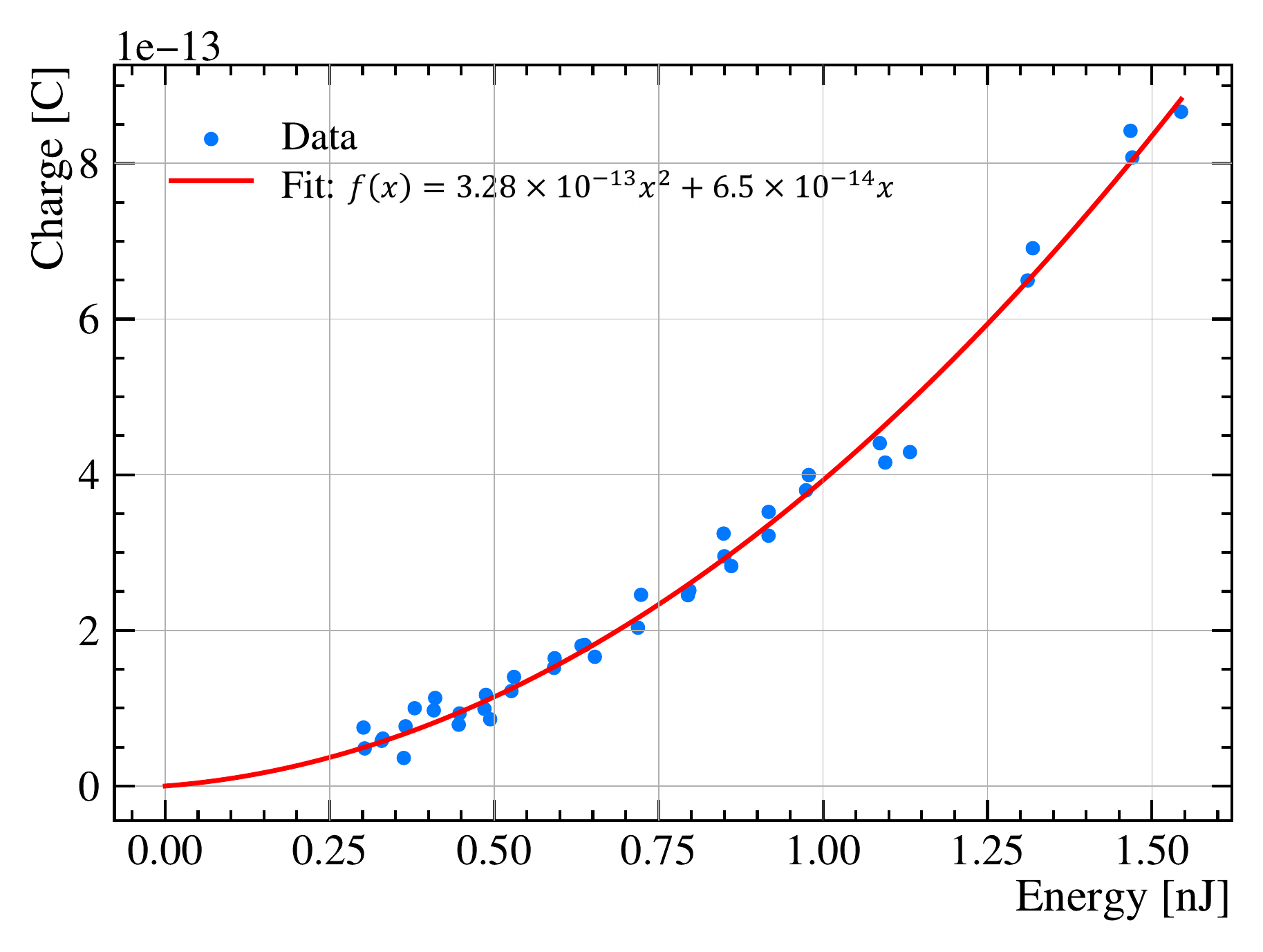}
    \caption{Charge vs energy per pulse for the diamond sample at a laser wavelength of 405~nm. A quadratic fit to the data points is shown.}
    \label{fig:DiodeChargeDepthScanExtraFromZero}
  \end{subfigure}
  \caption{Charge as a function of the pulse energy inferred from a calibrated SPA reference monitor. Both plots show a quadratic dependency, as expected from the TPA process. 
  }
  \label{fig:CiS_Quadratic_dep}
\end{figure}

To perform the energy scan measurements, the sample is mounted on the microscope stage and the $x$, $y$ and $z$ coordinates for four different corner points of the device are measured to ensure effective positioning of the DUT. The focal point of the laser is determined and measured. The pulse frequency is set to 10 kHz. The z-value and bias voltage are set to a constant value throughout the run. Data was collected for more than one full rotation of the NDF wheel in steps of one degree. The charge is calculated for each step of the NDF and the NDF angle is used to define the pulse energy using the correlation factor determined from the power calibration measurement. Fig.~\ref{fig:CiS_Quadratic_dep} shows the charge as a function of the pulse energy. This shows the expected quadratic relationship of charge and pulse energy according to theoretical predictions for both silicon and diamond (shown in Equation~\ref{eqn:CERN_model_TPA_charge})~\cite{TPAchargeDepend}.
\begin{equation}
    N_{\text{TPA}}(z) = \frac{E_{p}^{2}\beta_{2} \sqrt{\ln4}}{c\tau\hbar\pi^{\frac{3}{2}}}\left[\text{arctan}\left(\frac{d - (z -  z_{\rm {off}})}{z_{0}}\right) + \text{arctan}\left(\frac{z - z_{\rm {off}}}{z_{0}}\right)\right],
    \label{eqn:CERN_model_TPA_charge}
\end{equation}
where $N_{\text{TPA}}(z)$ is the TPA charge carrier density, $E_{p}$ is pulse energy, $\beta_{2}$ is the TPA coefficient, $\tau$ is the pulse duration, $d$ is the device thickness, $c$ is the speed of light, $z_{0}$ is the Rayleigh length, $z_{\rm {off}}$ is the $z$-axis offset and $\hbar$ is the Planck constant.

\begin{figure}%
    \centering
    \begin{subfigure}{0.49\linewidth}
    \includegraphics[width=\linewidth]{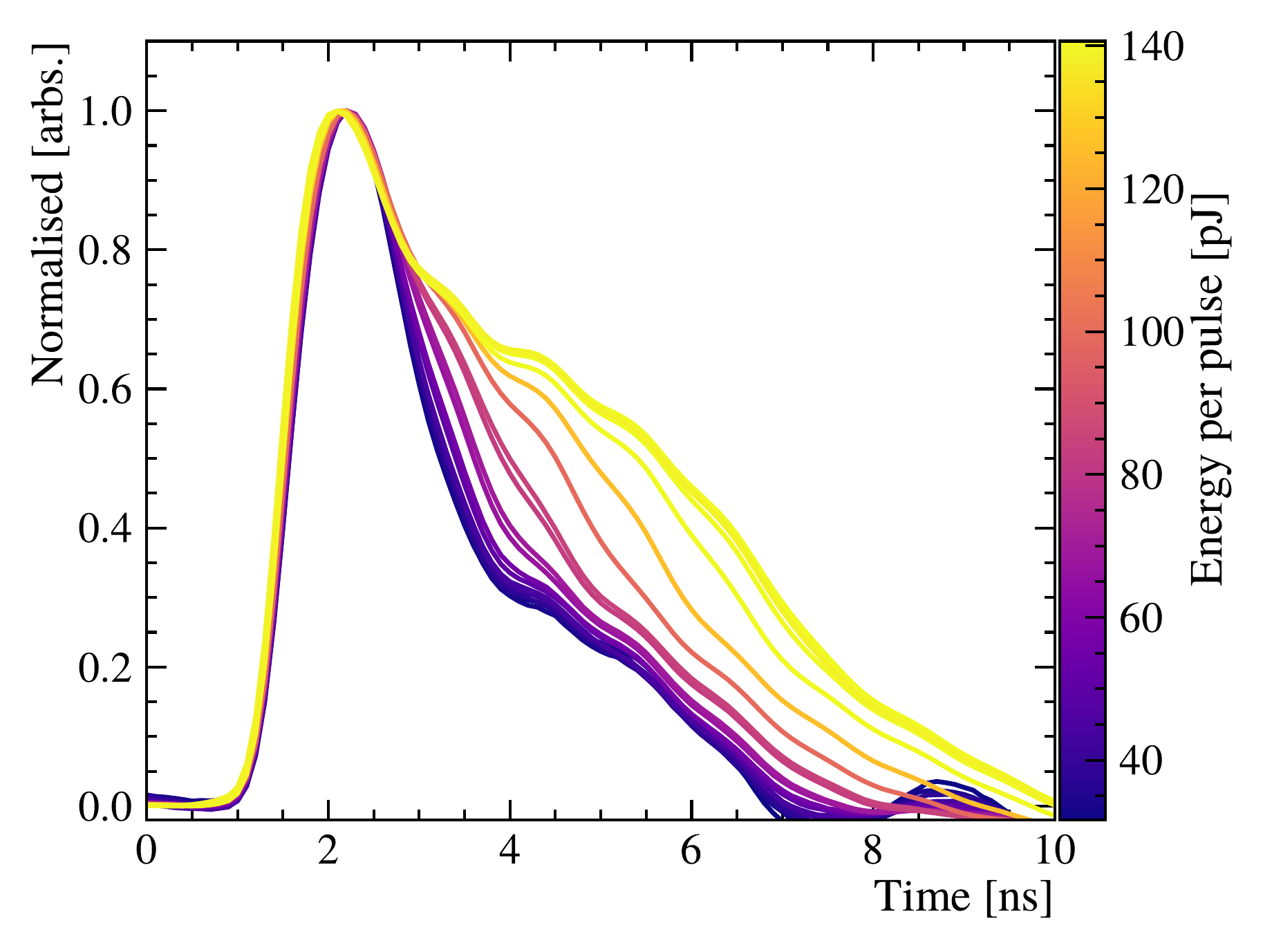}
    \caption{Signal waveforms for different pulse energies.} \label{fig:EnergyScanWaveforms}
  \end{subfigure}
   \begin{subfigure}{0.49\linewidth}
    \includegraphics[width=\linewidth]{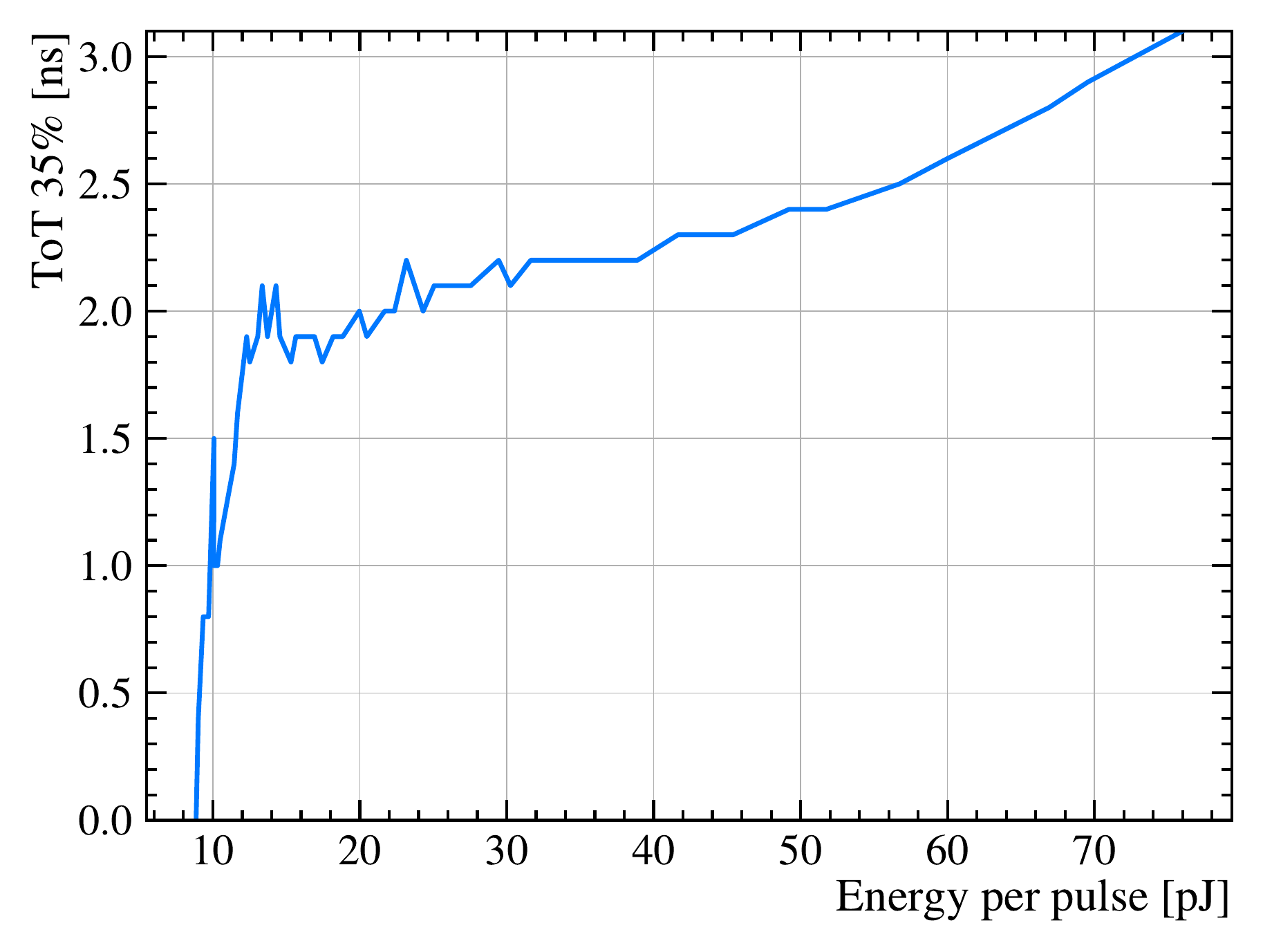}
    \caption{ToT (35\%) for different NDF steps.}
    \label{fig:ToT}
  \end{subfigure}
  \caption{Signal waveforms obtained during an energy scan by varying the steps/angle of the NDF wheel while utilising the CIS diode and a wavelength of 1550~nm. Different steps/angles correspond to different energy levels which are indicated in Fig.~\ref{fig:CiS_Quadratic_dep}. The plasma effect region can be seen for values above 55~pJ.}
  \label{fig:EnergyScan}
\end{figure}

High charge carrier densities might create an undesirable electron-hole plasma that affects the signal collection time and shape~\cite{plasmaEffect}. Fig.~\ref{fig:EnergyScanWaveforms} shows signal waveforms for a range of energy per pulse and it indicates that the collection time increases above a certain threshold as expected for the plasma effect. This effect is visualised by plotting the Time-over-threshold (TOT) of the signal as a function of the energy per pulse in Fig.~\ref{fig:ToT}. The threshold is set to 35\% of the peak signal, with a background subtraction of 0.005~V. Three different regions are identified: low $\mathrm{E}_\mathrm{p}$ (6 - 15~pJ), where measurements are background dominated, medium $\mathrm{E}_\mathrm{p}$ (16 - 55~pJ), which is the good S/B ratio region, and high $\mathrm{E}_\mathrm{p}$ (55~pJ and above), which is the plasma effect region. After this calibration stage, the NDF wheel angle is chosen in such a way it lies in the middle of the medium $\mathrm{E}_\mathrm{p}$ range

\FloatBarrier

\section{Depth Profiling and Beam Waist Characterisation}
\begin{figure}[h!]
    \centering
    \includegraphics[width=100mm,scale=0.5]{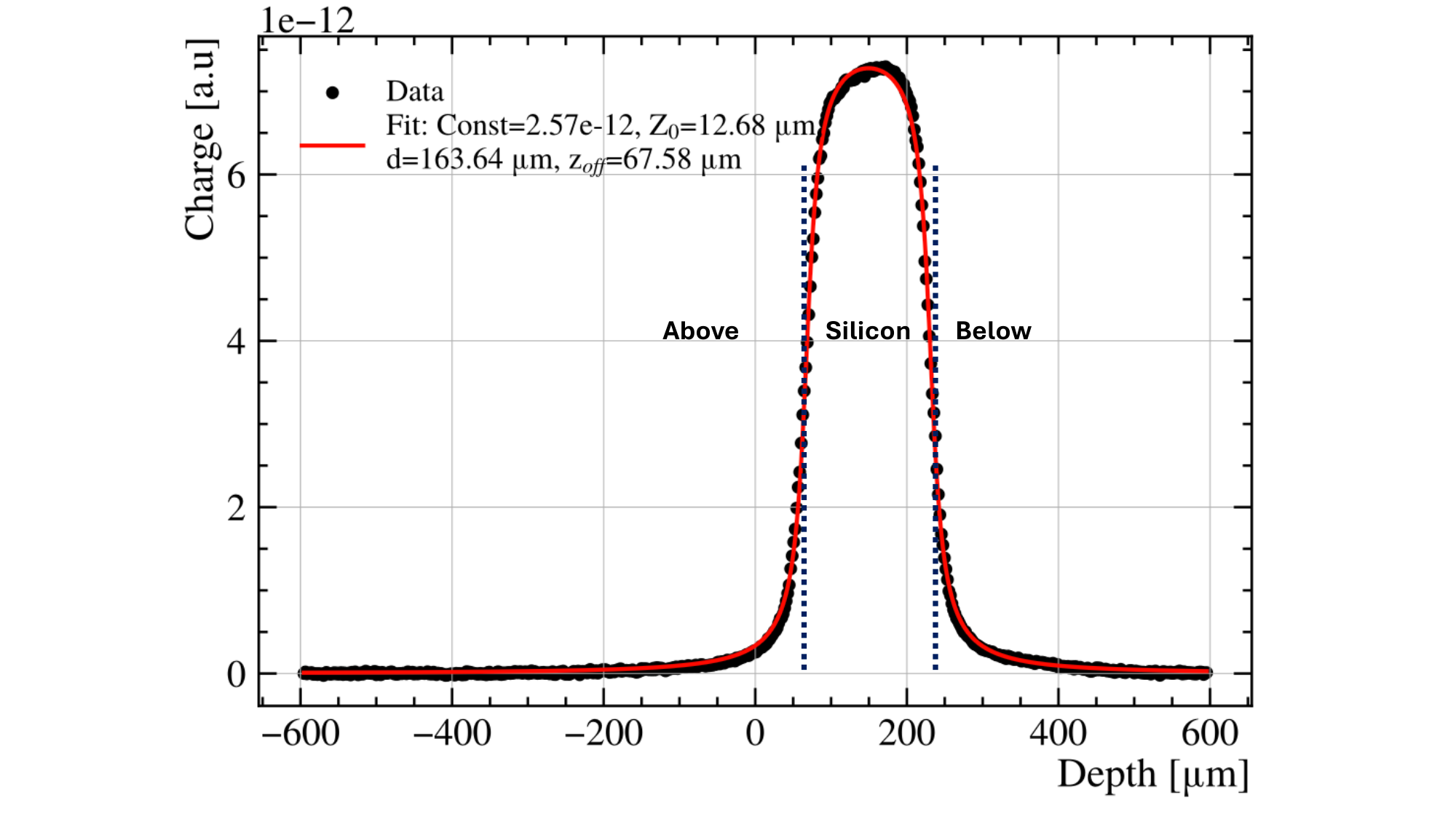}
    \caption{Corrected charge as a function of depth from the CIS diode at a laser wavelength of 1550~nm and a bias voltage of 130~V. 
    The data was fitted using the expression given in Equation~\ref{eqn:CERN_model_TPA_charge}.}
    \label{fig:CIS_charge_Depth}
\end{figure}

\noindent The TPA process has two main identifying features namely the quadratic dependence of the DUT signal on the laser power (as seen above) and the signal dependence on the depth scan. Since the amount of absorbed light is reliant on the position of the focal point, no signal is expected when the focal point is moved far outside in depth, the $z$-direction, of the sample. The laser is injected from the top of the device and the DUT is moved along the beam axis. The focal point is moved all the way through the device and outside of the device both above and below. Measurements were carried out by averaging over 1000 waveforms on the oscilloscope, at a laser pulse frequency of 10~kHz. For each $z$-value, the charge is calculated by integrating the DUT signal and accounting for the gain and resistance of the amplifier. Fig.~\ref{fig:CIS_charge_Depth} shows the generated charge as a function of $z$ with the depth scaled to correct for the effect of refraction within the device.  
The upper part of the device is at low values of $z$.  The charge is corrected by the square of the SPA monitor signal maximum value so that the mean observed charge is maintained. No charge is generated for TPA focussing outside of the device as there is a negligible single photon absorption cross-section at the laser wavelength used. This therefore is another evidence of the TPA process.

\FloatBarrier

The beam waist is estimated using the knife-edge technique~\cite{NIKHEFTPA}. By moving the voxel focal point across a metallisation barrier at different depths, it is possible to infer the beam waist (width) and, therefore, measure the voxel perpendicular resolution.

\begin{figure}[h!]
    \centering
    \includegraphics[width=0.8\linewidth, trim={0cm, 4cm, 0cm, 2cm}, clip]{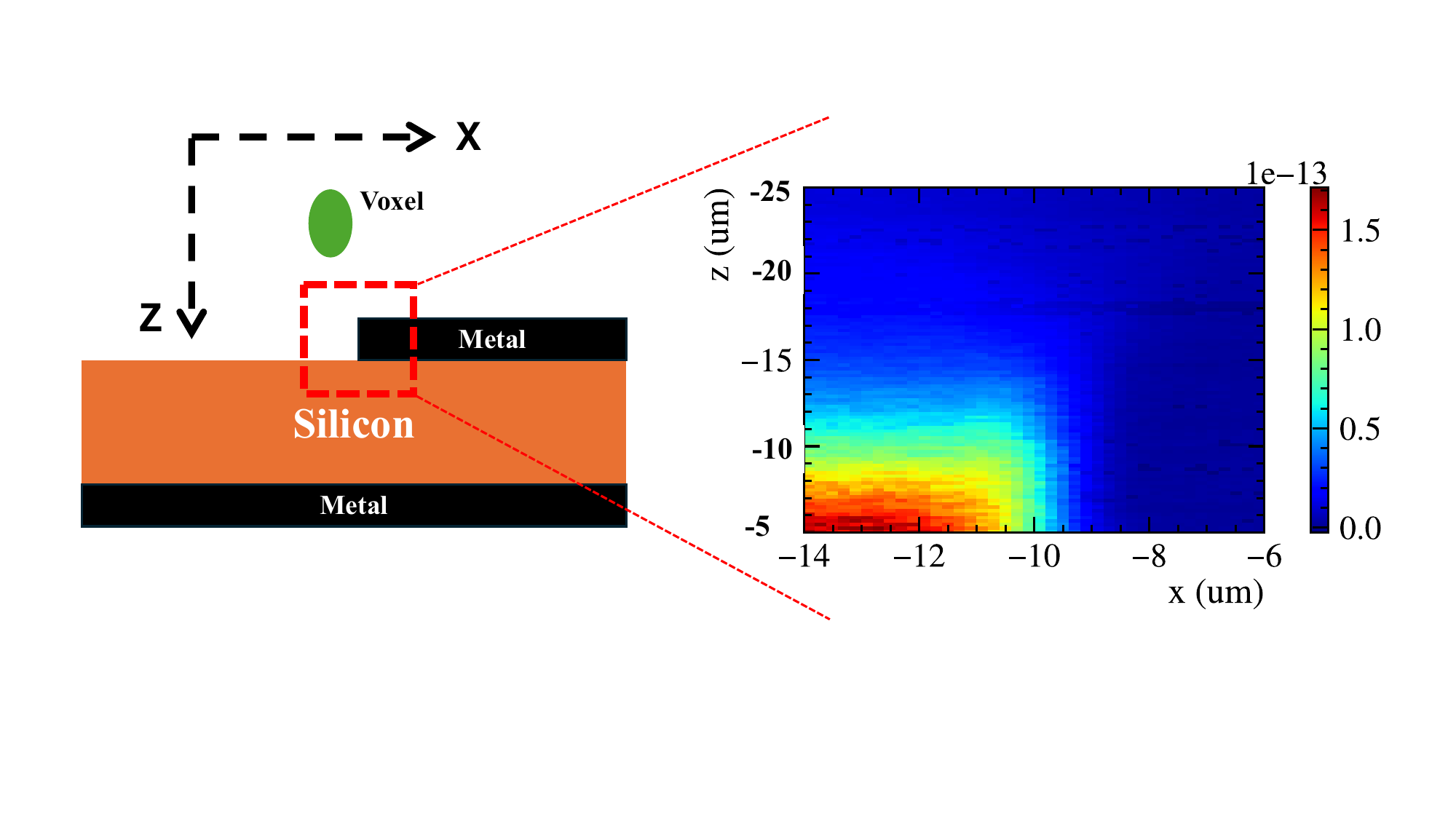}
    \caption[A schematic of a knife edge scan.]{A schematic of a knife edge scan. The schematic of the DUT with a metallization on the top is shown on the {\em left} and a plot of the charge generated at the knife edge region is shown on the {\em right}.}
    \label{fig:Knife_edge_schematic}
\end{figure}

The data presented was taken with the CIS sensor at a constant reverse bias voltage of 150~V. This is well above the full depletion voltage of the device. Step sizes of 0.2~$\upmu$m were taken in both the x and z directions to ensure effective sampling of the beam profile when the metal edge is in focus, as shown in Fig.~\ref{fig:Knife_edge_schematic}. The measurements were averaged over 1000 waveforms and the pulse frequency adjusted to 10~kHz. 

\begin{figure}[h!]
    \centering
    \includegraphics[width=100mm,scale=0.5]{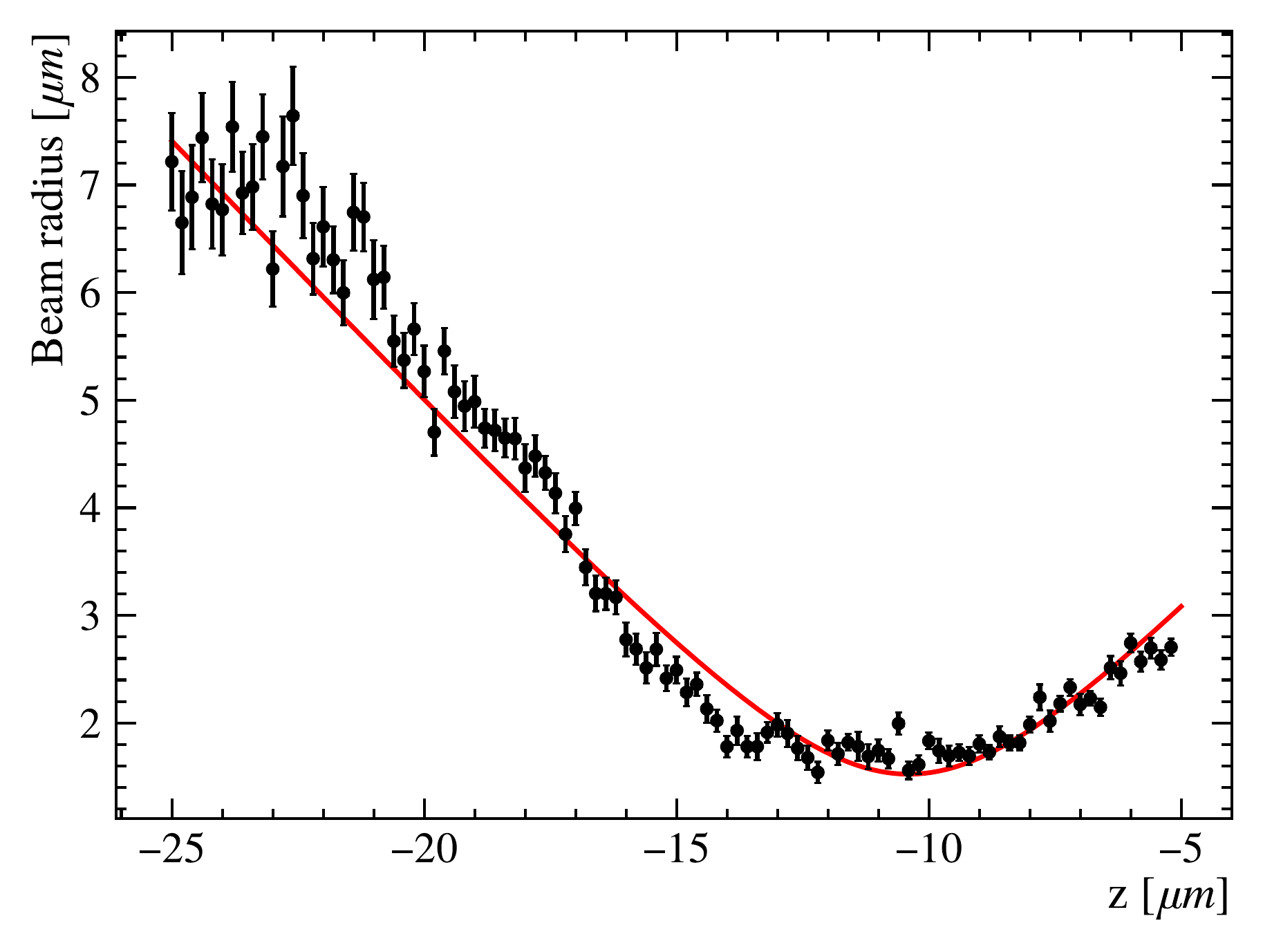}
    \caption{Beam radius as a function of depth (z) for the diode. The values of the Rayleigh length in silicon $z_{0}$ = $12.20 \pm 0.47$~$\upmu$m and the beam radius at the waist $w_{0}$ = $1.53 \pm 0.04$~$\upmu$m were determined, corresponding to the voxel height and width respectively.}
    \label{fig:Beam Waist}
\end{figure}

The measurements were averaged over 1000 waveforms and the pulse frequency adjusted to 10~kHz. To calculate the beam waist, $w(z)$, the charge profile in the x-direction is fitted with Equation~\ref{eqn:erf} for every value of z~\cite{NIKHEFTPA}.
\begin{equation}
    \centering
    f(x) = \frac{C}{w^{2}(z)}\left[1 + {\rm erf}\left(\frac{2(x - x_{{\rm off}})}{w(z)}\right)\right] ,
    \label{eqn:erf}
\end{equation}
where the fit parameters are; C - a constant, x$_{off}$ - offset in the x-coordinate and $w(z)$ - beam radius. The behaviour of the beam radius as function of the depth can be predicted by the equation below~\cite{MoritzTheses}:

\begin{equation}
    \centering
    w(z') = w_{0}\sqrt{1 + \left(\frac{z'}{z_{0}}\right)^{2}},
    \label{eqn:radius}
\end{equation}

where $w_0$ corresponds to the minimum waist and $z'$ is the distance from the focal plane along the $z$-axis. By fitting this equation, the minimum waist was obtained, as shown in Fig.~\ref{fig:Beam Waist} the voxel dimensions were estimated to be $12.20 \pm 0.47$~$\upmu$m longitudinally and $1.53 \pm 0.04$~$\upmu$m perpendicularly.

Considering silicon as medium and a lens with NA=0.5, the theoretical expectations for $w_{0}$ and $z_{0}$ are 0.99~$\upmu$m and 6.86~$\upmu$m respectively. However, due to the optical distortions of a `real' lens, the measured parameters are significantly larger as typically observed. Chromatic effects are definitely present, but they have been neglected in other studies. Furthermore, no clear indications of sizeable spherical aberrations were observed, as no significant depth dependency was observed on the Silicon diode~\cite{Table_setup1}.

\FloatBarrier

\section{Reflection Modelling}
Understanding the effects of optical reflections is essential, as they are often unavoidable in the characterisation of devices. Some sensors feature backside surface metallisation, where materials with refractive indices different from silicon introduce reflection at the interface. The extent of reflection depends on the degree of refractive index mismatch and the proximity of the laser focal point to the backside surface.
Three reflection models were therefore developed to further understand these effects of reflection. The sensor used in this section is the CNM diode which has an aluminium coated back surface. 

The first model (\textit{Convolution model}) was constructed by taking into account the voxel size dimensions in the induced charge. For a truly point like voxel, the expected charge response by depth scan would be a rectangular function given by the thickness of the device, assuming no optical distortions. Since the voxel dimensions are not negligible, the response from a sample without reflection on the back surface could be simply represented by the convolution of the rectangular function  and the voxel. The fully reflected component can similarly also be easily represented  with the amplitude of the reflected charge allowed to be different due to the reflection coefficient of the back of the sample. Close to the reflection interface, there is an additional contribution from the overlap between the direct and reflected pulse parts. This contribution is also modelled in the same way. The boundaries of these three regions were defined empirically and their rectangular function amplitudes were fitted to the data after the voxel convolution. 

The second model is the \textit{3D reflection model}. This involves calculating the carrier density contributions: $n^D$ from the direct voxel, $n^R$ from the reflected voxel, and $n'$ from the superposition terms. The model includes an effective parameter NA$_{\rm{eff}}$ to account for the aberration of the voxel shape caused by refraction. Subsequently, the two-photon absorption carrier density, $n_{\rm{TPA}}$ is determined as a function of the spatial coordinates $(z,r)$ (Equation~\ref{eqn:reflect_ntpa_3D_model}). This model also incorporates the delay time, $\Delta t$ introduced by the difference in optical path between the direct and reflected voxel. 
\begin{equation}
    \begin{split}
    n_{\rm{TPA}}(z,H,r) & =\frac{\beta_2}{2\hbar\omega}\int_{-\infty}^\infty \rm{d\it t}\ \it I^2(z,r,t)  \\
    & = \frac{\beta_2\tau}{4\hbar\omega}\sqrt{\frac{\pi}{2\ln{2}}}I_{\rm D}^{2}(z-H,r) + \frac{\beta_2R^{2}\tau}{4\hbar\omega}\sqrt{\frac{\pi}{2\ln{2}}}I_{\rm D}^{2}(z+H,r) \\
    & + \frac{\beta_2R\tau}{2\hbar\omega}\sqrt{\frac{\pi}{2\ln{2}}}I_{\rm D}(z+H,r)I_{\rm D}(z-H,r)e^{-\frac{2\ln 2(\Delta t)^{2}}{\tau^2}},
    \end{split}
    \label{eqn:reflect_ntpa_3D_model}
\end{equation}
where $R$ is the reflectance of the back surface of the sample, $H$ is the centre position for the direct voxel, $I$ is the total intensity and $I_{\rm D}$ is the total intensity of the direct voxel.

The third description of the reflection uses the model detailed in  Ref.~\cite{pape2024characterisation} (\textit{CERN model}). The model is based on the assumption that the duration of the laser pulse is negligible, i.e. that the illumination is instantaneous. The model takes into account the direct component of the voxel, the voxel reflection, and the interaction of the direct and reflection components. The interaction is modelled using the surface correction function that controls the reflection onset, a weighting function that peaks where the overlap is strongest, and the non-interacting contributions from both the direct and reflected components.

\begin{figure}[h!]
    \centering
    \includegraphics[width=0.99\linewidth]{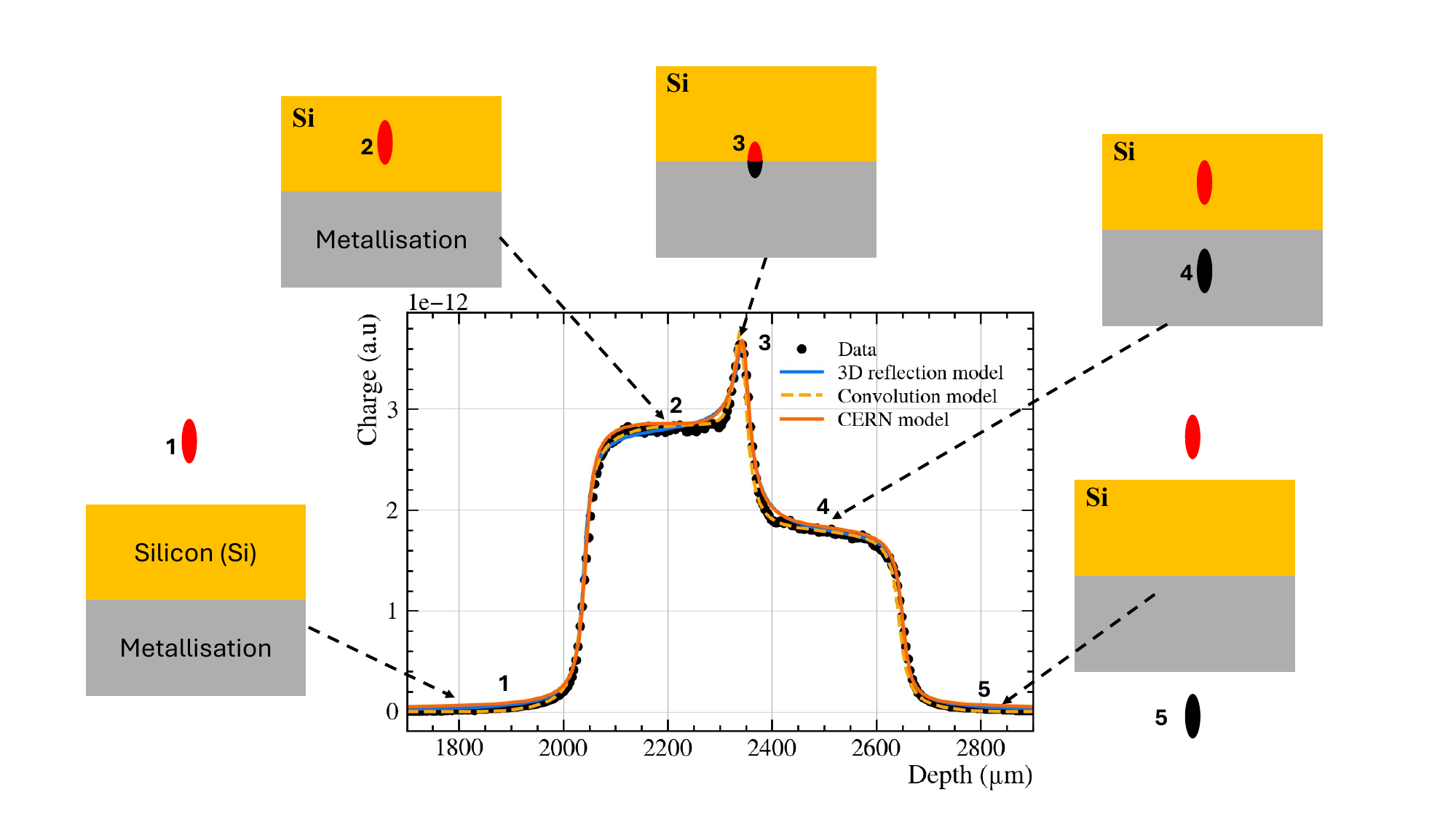}
    \caption{ 
    Charge with depth scan experimental data obtained using the CNM PIN diode. A comparison is made with a fit using convolution, 3D reflection and CERN models (see text). The plot is annotated with laser focus positions numbered 1 to 5.  The black spot shows where the laser focus would have been if there was no reflection. The red spot shows the effective voxel position.}
    \label{fig:TPAconvolveReflection}
\end{figure}
Fig.~\ref{fig:TPAconvolveReflection} shows how well each model matches with the depth scan data. All models describe the general function of the data well. The reflectance of the material on the back side surface was found to be compatible in all models: 80.0 $\pm$ 0.6\% for the convolution model, 79.1 $\pm$ 0.7\% for the 3D model and 71.0 $\pm$ 14.1\% for the CERN model. Applying a Kolmogorov-Smirnov (KS) test \cite{lilliefors1967kolmogorov} and Root mean Square Error (RMSE) statistic \cite{hyndman2006another} the 3D reflection model was found to be the most consistent with the experimental data.
The 3D reflection model is the only model to incorporate the three dimensional determination of the charge and to account for the aberration of the voxel shape caused by refraction.
\FloatBarrier

\section{Two-Photon Absorption Simulation}
\label{subsec:simulation}
The simulation was performed using KDetSim software~\cite{KDetSim}. KDetSim was optimized to simulate TCT signals and it is integrated with the ROOT framework~\cite{ROOTCERN}. In KDetSim, the signal is considered to be distributed along a line in segments (buckets). Since the radial resolution is much better than the resolution over depth, the one dimension projection of the voxel over the depth axis is considered for the modelling.
The modified numerical aperture used in the simulation was estimated to be 0.37 based on the Rayleigh length obtained by the charge by depth fit in Fig.~\ref{fig:CIS_charge_Depth}. 

\begin{figure}[h]
    \centering
    \includegraphics[width=0.49\linewidth]{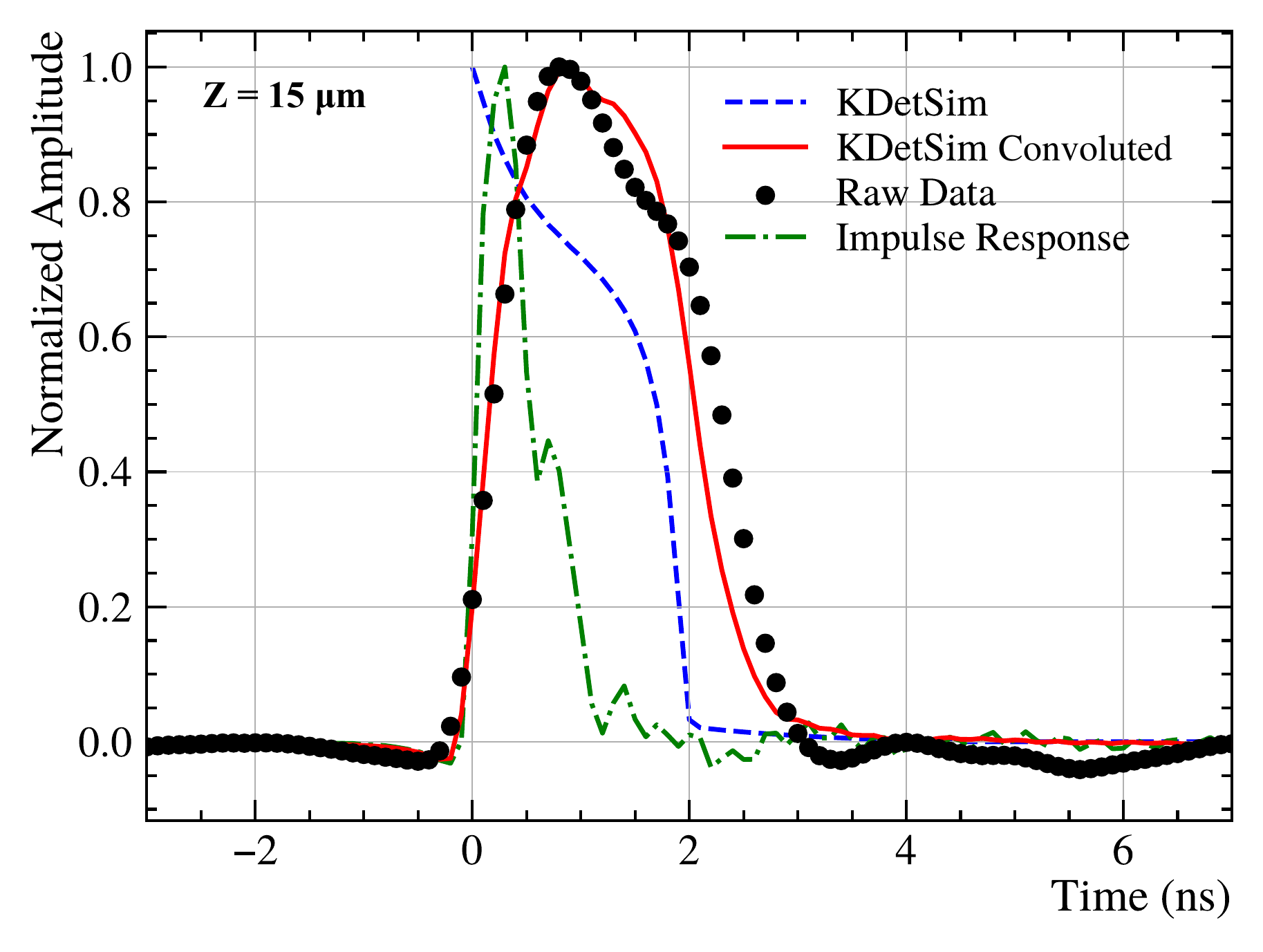}
    \includegraphics[width=0.49\linewidth]{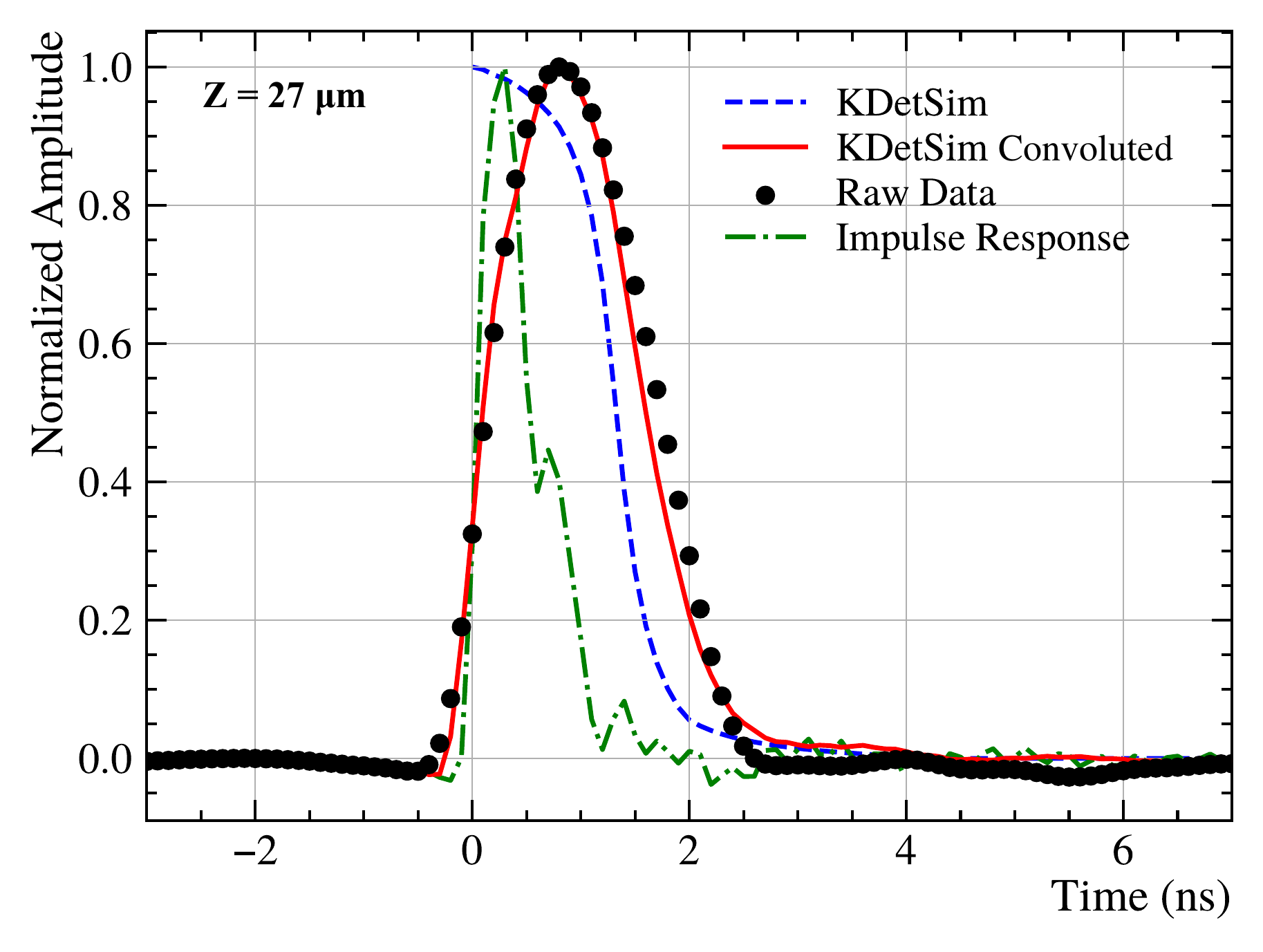}
    \includegraphics[width=0.49\linewidth]{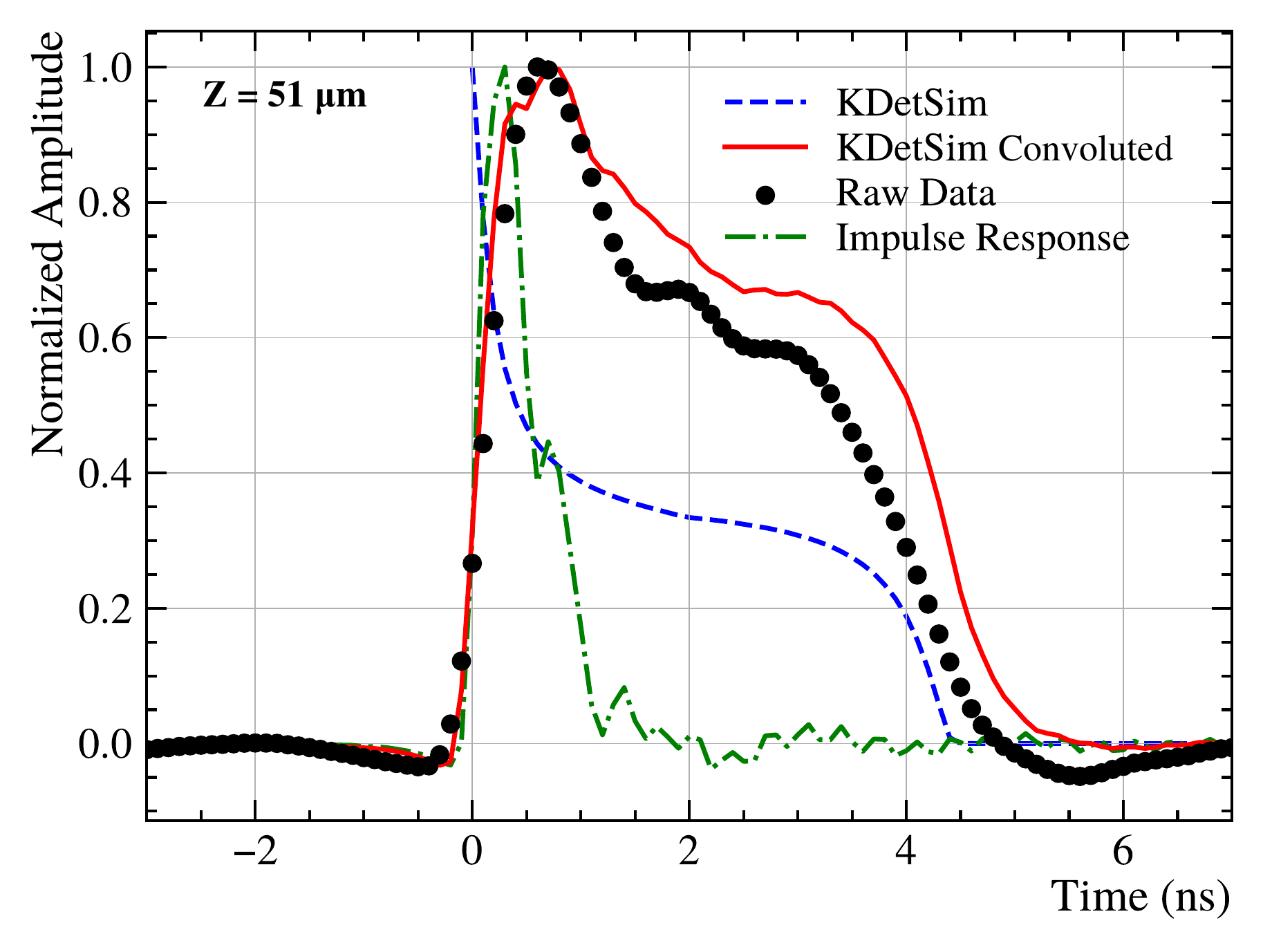}
    \includegraphics[width=0.49\linewidth]{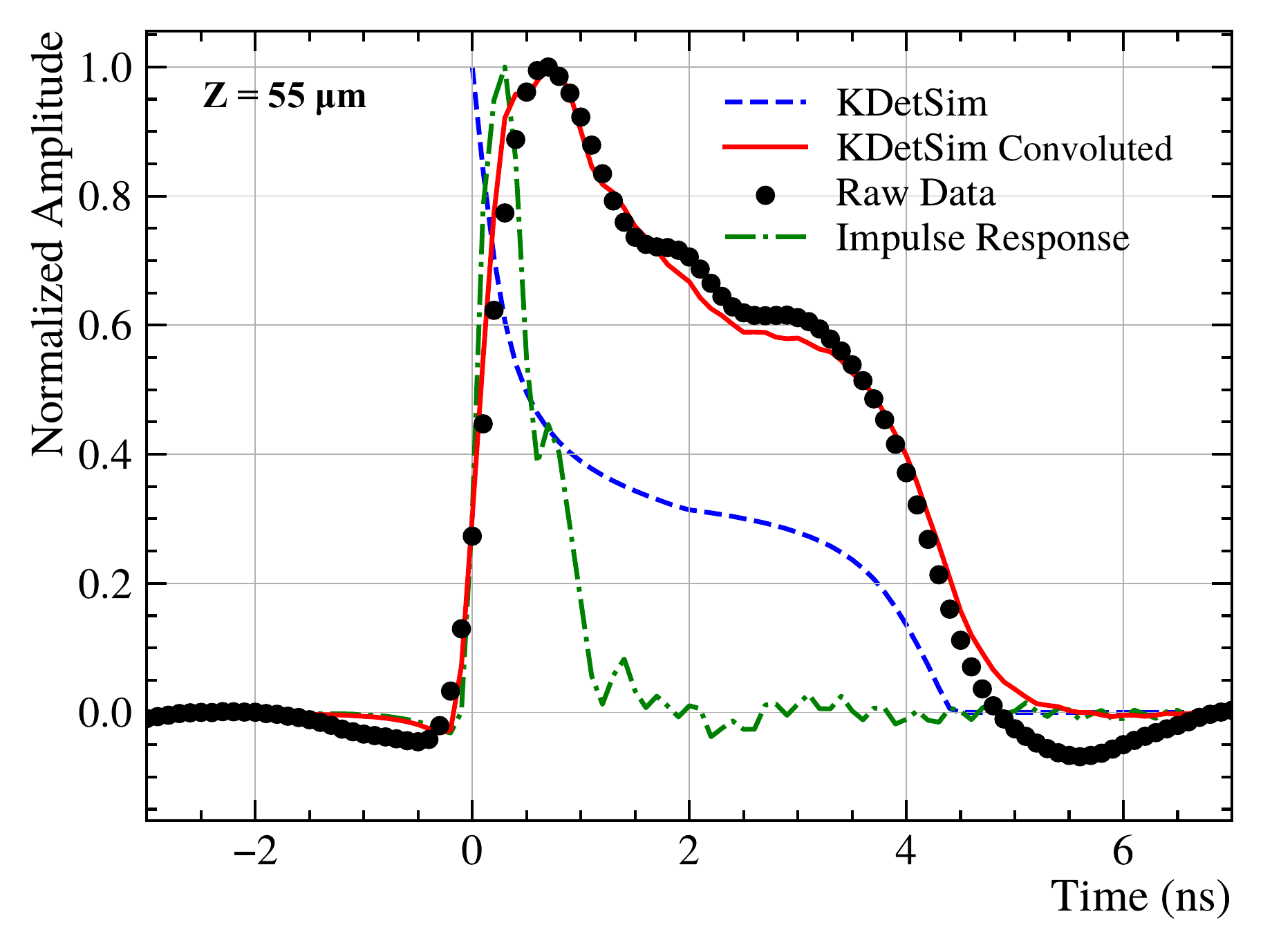}
    \caption{CIS diode signal waveform and the prediction of the KDetSim simulation convoluted with the amplifier response. The waveforms are given with the laser focus at the device air-silicon interface (top left, $z=15$~$\upmu$m), at the middle of the device (top right, $z=27$~$\upmu$m), at the backside (bottom left, $z=51$~$\upmu$m) and set beyond the device thickness (bottom right, $z=55$~$\upmu$m). These {\em z}-values are those in the lab system, not accounting for the refractive index correction.}
    \label{fig:KDetDepthTimeConvolution}
\end{figure}

\begin{figure}[h]
    \centering
    \includegraphics[width=0.49\linewidth]{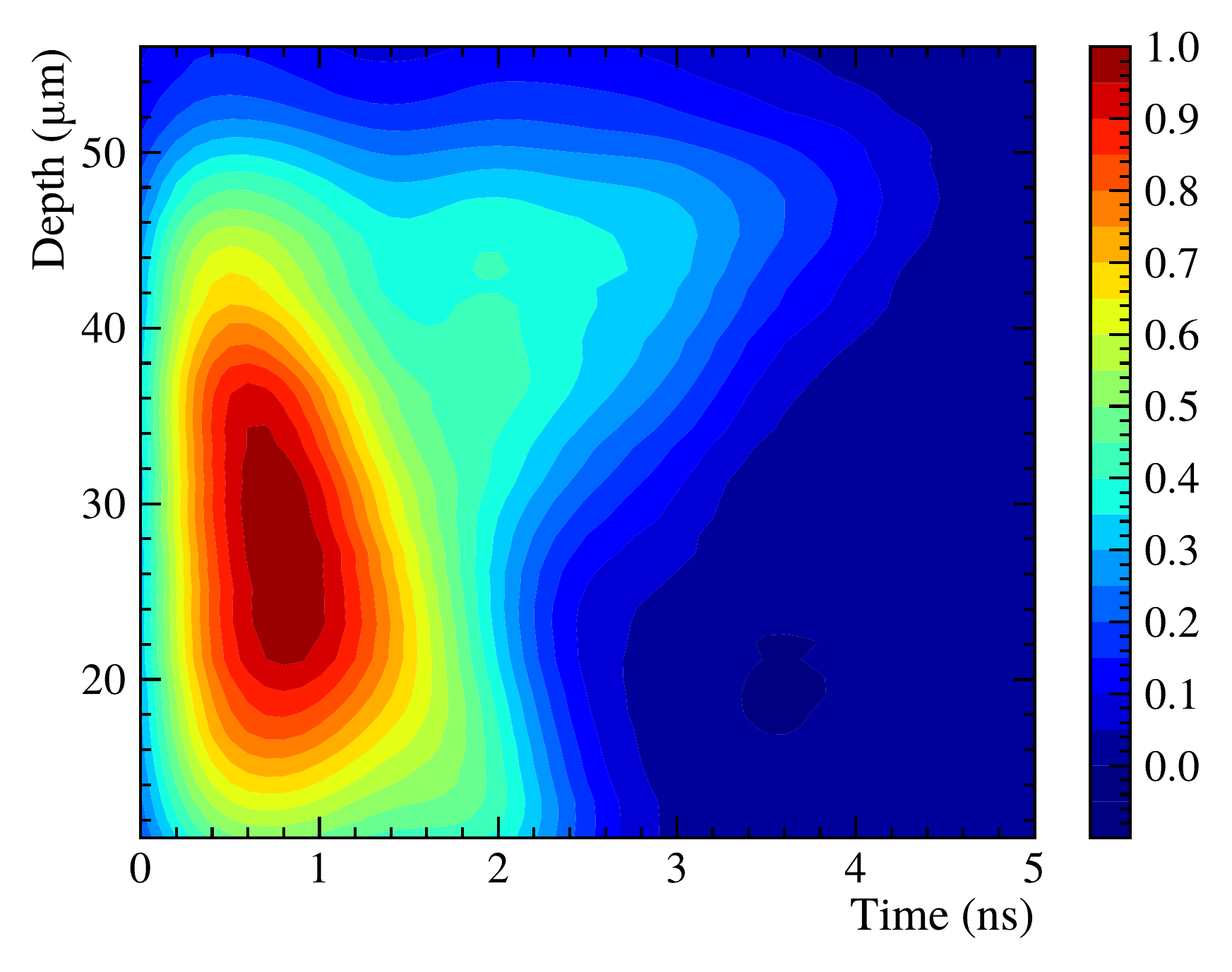}
    \includegraphics[width=0.49\linewidth]{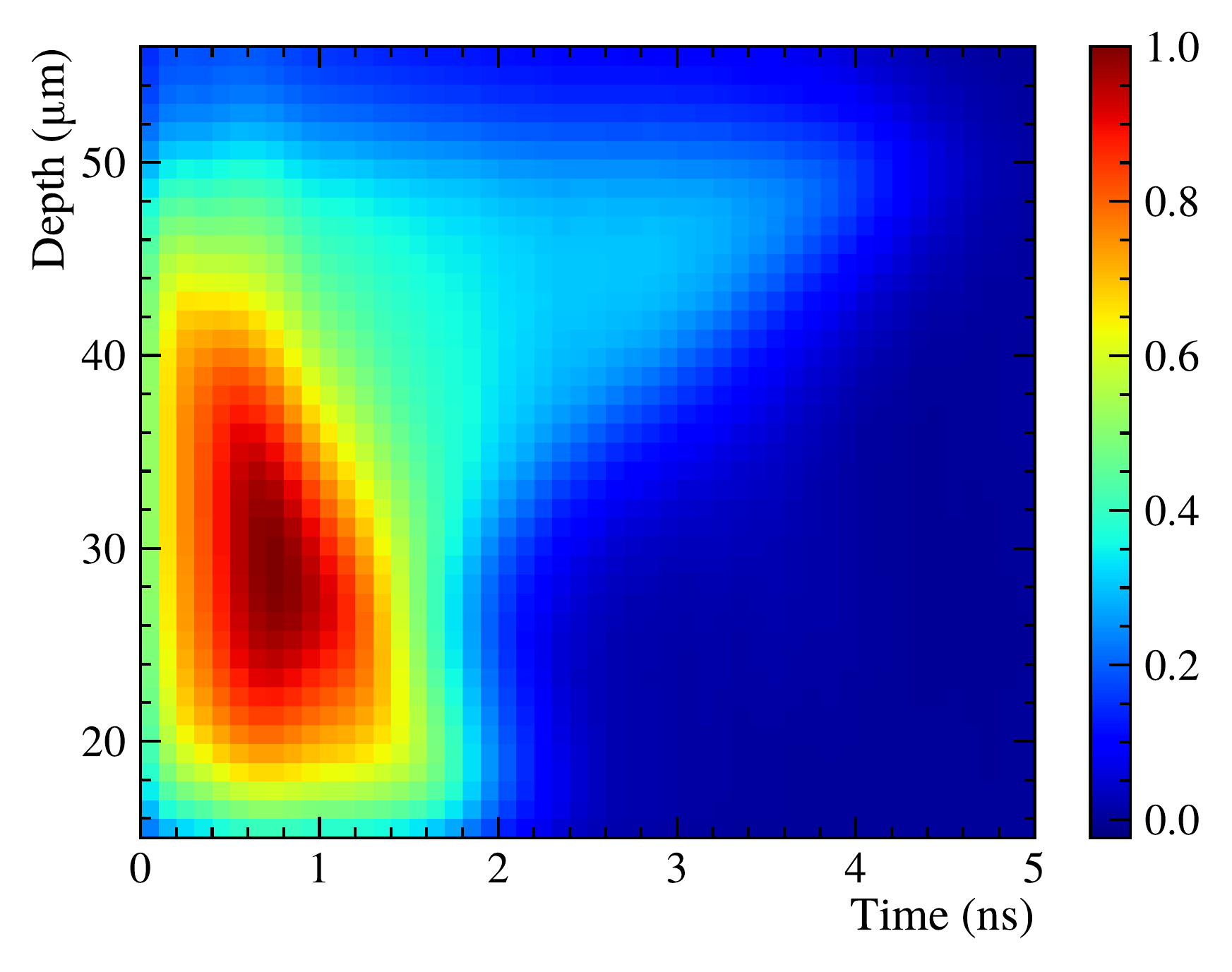}
    \caption{A 2D depth-time scan of TPA signal for the CIS diode from the experiment (left figure) and using the KDetSim simulation convoluted with the amplifier response (right figure).}
    \label{fig:KDetSim_depth_scan_conv_experiment}
\end{figure}

To compare the simulation and measurement, the signal waveform must be aligned spatially (depth) and in time. The spatial alignment was done based on the rising/falling edge of the charge as a function of depth at a constant fraction of 50\%. The time alignment was performed by using a constant fraction discriminator at 30\% and 70\% of the signal maximum to predict the starting time of the waveform. The amplifier response function was used to convolute the simulation in order to describe the expected distortion from the electronics before comparing with the measurements. The measured CIS diode signal waveforms and the prediction of the simulation are shown in Fig.~\ref{fig:KDetDepthTimeConvolution}. The  2D depth-time scan of the experiment and in the simulation are shown in Fig.~\ref{fig:KDetSim_depth_scan_conv_experiment}. 

Overall, the simulation qualitatively reproduces the main features observed in the data. The signal maxima of both the measurement and simulation are well aligned in time. When considering the full width half maximum (FWHM), a percentage difference of up to 20\% between the simulation and the data is observed at the backside of the sample, in the shoulder region of the distribution along the time axis, as shown in the bottom left of Fig.~\ref{fig:KDetDepthTimeConvolution}.

\FloatBarrier

\section{Conclusion}

The two-photon absorption process provides an excellent method for the 3D characterization of solid-state sensors response. The PHAROS coupled with an OPA system (ORPHEUS) provides a very flexible system capable of reaching the appropriate wavelength range for the two-photon absorption process in several substrates including silicon and diamond which are demonstrated in this paper. This wavelength flexibility coupled with an automatized data taking system provides an excellent platform for the characterisation of several solid-state sensors. The set-up is simple to use, faster and far more affordable than a test-beam setup, while providing detailed characterisation information.

The system was commissioned and the TPA process was established based on the results obtained with the quadratic behaviour of the charge induced in a silicon diode and a diamond planar sensor. In addition, the charge induced in the silicon diode as function of depth also matches the expectation for the two-photon absorption process. The voxel characterisation shows a Rayleigh length of $12.20 \pm 0.47$~$\upmu$m and the beam radius at the waist of 1.53 $\pm$ 0.04~$\upmu$m. A diode with metallisation on the backside was also used to investigate the reflection on the back of the sample. Three models were fitted to the data and the best agreement was obtained with the 3D reflection model that incorporated the three dimensional determination of charge, the time delay between the direct and reflected voxel, and accounted for the aberration of the voxel shape caused by refraction.

Lastly, the signal waveforms obtained for a silicon diode were successfully modelled using a simplified approach in KDetSim including the convolution with the amplifier response. The main features of the charge response in depth and time were reproduced. The signal maxima of both the measurement and simulation are aligned in time. Percentage differences of up to 20\%, along the time axis, between data and simulation were observed when considering the FWHM and attributed to the modelling of reflection from the back surface. 

\section{Acknowledgements}
We would like to acknowledge the contributions by Neil Moffat of CNM for providing the CNM PIN diode sample which was used for the reflection measurements. We thank the former RD50 Collaboration, now DRD3, for the CIS diode sample. We are grateful to the RD42 collaboration for the diamond sample. We thank the CERN TPA team for extremely fruitful discussions. Thanks also go to the CERN team for characterising the amplifier response that is used in this paper. Part of this project was funded by UKRI/STFC grants ST/V003410/1, ST/Y005457/1, ST/W000601/1, MR/T021519/1 Fellowship, and the laser was funded by EP/V036343/1 grant. The work was also funded by the Commonwealth Scholarship Commission and the Foreign, Commonwealth and Development Office of the UK. We are grateful for their support. All views expressed here are those of the authors not the funding body.

\bibliographystyle{JHEP}
\bibliography{biblio.bib}






\end{document}